\documentstyle[12pt]{article}
\newcommand{\eq}{\begin{equation}}
\newcommand{\en}{\end{equation}}
\newcommand{\eqa}{\begin{eqnarray}}
\newcommand{\ena}{\end{eqnarray}}
\newcommand{\eqs}{\begin{displaymath}}
\newcommand{\ens}{\end{displaymath}}
\newcommand{\eqas}{\begin{eqnarray*}}
\newcommand{\enas}{\end{eqnarray*}}

\advance\oddsidemargin by -\textwidth
\advance\oddsidemargin by -1in
\evensidemargin=\oddsidemargin
\begin{document}

$\mbox{ }$
\vspace{-3cm}
\begin{flushright}
\begin{tabular}{l}
{\bf KEK-TH-411 }\\
{\bf KEK preprint 94 }\\
August 1994
\end{tabular}
\end{flushright}

\baselineskip18pt
\vspace{1cm}
\begin{center}
\Large
{\baselineskip26pt \bf A Note on String Field Theory\\
                       in the Temporal Gauge}\footnote{
Based on a talk presented by N.I.
in the workshop {\it Quantum Field Theory, Integrable Models and Beyond},
Yukawa
Institute for Theoretical Physics, Kyoto University, 14-18 February 1994.}
\end{center}
\vspace{1cm}
\begin{center}
\large
$\mbox{{\sc M. Ikehara,}}^{\star ,\dagger }~
\mbox{{\sc N. Ishibashi,}}^{\star}~
\mbox{{\sc H. Kawai,}}^{\star}$\\
$\mbox{{\sc T. Mogami,}}^{\star ,\ddagger}~
\mbox{{\sc R. Nakayama}}^{\diamond }~
\mbox{{\sc and}}~
\mbox{{\sc N. Sasakura}}^{\star}
$
\end{center}
\normalsize
\begin{center}
$~^\star$ {\it KEK Theory Group, Tsukuba, Ibaraki 305, Japan}\\
$~^\dagger$ {\it Department of Physics, University of Tokyo, Bunkyo-ku,
Tokyo 113, Japan}\\
$~^\ddagger$
{\it Department of Physics, Kyoto University, Kitashirakawa, Kyoto 606,
Japan}\\
$~^\diamond$ {\it Department of Physics, Hokkaido University, Sapporo 060,
Japan}
\end{center}
\vspace{2cm}
\begin{center}
\normalsize
ABSTRACT
\end{center}
{\rightskip=2pc 
\leftskip=2pc 
\normalsize
In this note, we review the recent developments in the string field theory
in the temporal gauge.
\vglue 0.6cm}

\newpage
\section{Introduction}
\hspace{5mm}
String theory yields the most promising theory of quantum gravity.
However, a
nonperturbative treatment of string theory is indispensable for relating it
to the phenomena we see in experiments.
For example, string theory possesses innumerable classical vacua. People
believe that one of them (or one superposition of them) is selected by
a stringy nonperturbative effect.
In order to treat stringy nonperturbative effects, first-quantized approach of
string theory is inadequate and string field theory\cite{SFT} is necessary.
Especially we need a formulation of string field theory in which
a nonperturbative treatment is possible as the lattice approach was in the
QCD case.

A string field theory corresponds to a rule to
cut the string worldsheets into vertices and propagators, or in other words, a
way to fix the reparametrization invariance.
Recently a new kind of string field theories are
proposed for $c=1-\frac{6}{m(m+1)},~m=2,3,\cdots$ noncritical string\cite{IK}
\cite{IKtwo}\cite{IIKMNS}.
It is based on a gauge fixing \cite{KKMW}
of the reparametrization invariance,
which can naturally be considered on dynamically triangulated worldsheets.
The gauge is called the temporal gauge and the string field theory thus
constructed is called the string field theory in the temporal gauge.
Since this gauge is natural on dynamically triangulated worldsheets,
it is conceivable that the string field theory is related to the matrix
model approach in which a nonperturbative treatment of noncritical string
theory is possible.
Indeed, in this kind of string field theory, one can obtain the
Schwinger-Dyson (S-D) equations,
which  yield nonperturbative results. Namely, the Virasoro and $W$
constraints\cite{FKN} can be derived from the
S-D equation and all the results of the matrix model are reproduced.
Therefore temporal gauge string field theory is a powerful tool to investigate
noncritical strings nonperturbatively.

Let us briefly explain how the temporal gauge is useful.
The definition of the time coordinate in
\cite{KKMW}\cite{IK} is as follows.
Suppose a two dimensional surface with boundaries.
One can define the time coordinate of a point as the
geodesic distance from the set of the boundary loops.
The choice of the time coordinate
partially fixes the general coordinate invariance. Here we will call
the gauge corresponding to such a time coordinate the temporal gauge.
In the ADM language, taking this time coordinate corresponds to fixing the
lapse function $N$ to be unity. Therefore
the metric looks like,
\eq
ds^2=dt^2+h(x,t)(dx+N^1(x,t)dt)^2,
\label{gauge}
\en
at least locally.
In \cite{FIKN}, two dimensional quantum gravity was studied by fixing the gauge
further as $\partial_xh=0$. Such a gauge was called the ``temporal gauge''.
In \cite{NAK}, another gauge fixing $N^1=0$, which was called the ``proper time
gauge'', was pursued.

In two dimensions, this ADM-type gauge choice is
not so common as the conformal gauge.
However, as was elucidated by the authors of \cite{KKMW}, such a definition of
time is rather natural in the framework of the dynamical triangulation.
They defined the operation called ``one-step deformation'' of a loop, which
exactly coincides with the discrete
evolution of the loop in the coordinate frame considered
here. Thus if one starts from a loop, one gets a loop one step inside by this
deformation (see Fig. 1).
In \cite{KKMW}, the authors showed that this one-step deformation
has a well-defined continuum limit in the pure gravity case.

Thus it is likely that the gauge in eq.(\ref{gauge}) is useful in interpreting
the dynamical triangulation techniques in the continuum language. Indeed
the S-D equations of the matrix models are closely connected
with this one-step deformation.
Suppose the partition functions corresponding to surfaces with boundaries.
The S-D equations of the
matrix models describe the change of the partition functions when one
takes a triangle away from a boundary. It is easy to see that the one-step
deformation of a boundary loop corresponds to taking all the triangles away
along the boundary. Therefore the result of the one-step deformation can be
described as a sum of the S-D equations. In the continuum limit, the former
is an integration of the latter along the boundary loops.
As will become clear later, the equations describing the results of the
one-step deformations are the S-D equations of the temporal gauge
string field theory.

Hence the temporal gauge string field theory is closely related to the matrix
model approach and is powerful in analysing noncritical string theory
nonperturbatively.
Therefore if the temporal gauge string field theory were constructed for the
critical string, it might be a useful tool to study the nonperturbative
effects of string theory.
In this note, we will review the results obtained so far about the
temporal gauge string field theory.

The organization of this note is as follows. In section 2, we will consider
the temporal gauge in a simpler example, i.e. particle case. We show that
quantization of particle field theory in the temporal gauge results in
the stochastic quantization of it. As a toy model, we can learn many lessons
about string theory from it. In section 3, we turn to $c=0$ string theory
and discuss it in the temporal gauge. We will construct the string field
Hamiltonian in the temporal gauge. In section 4, we will deal with
$c=1-\frac{6}{m(m+1)}$ string. Section 5 is devoted to summary and discussions.

\section{Particle Field Theory in the Temporal Gauge}
\hspace{5mm}
As a warm-up, we will construct the Hamiltonian of a particle field theory
in the temporal gauge. Suppose a Feynmann graph of the (Euclidean) $\phi^4$
theory
(see Fig.2) with the action
\eq
S[\varphi ]=\int dx^4(\frac{1}{2}
\partial_\mu \varphi \partial_\mu \varphi +\frac{m^2}{2}\varphi^2
+\frac{\lambda}{4} \varphi^4).
\en
By the temporal gauge, we here mean the following way to introduce a time
coordinate on
the Feynmann graph. The time coordinate of a point on the graph is defined
to be the proper time from the nearest boundary (or the starting point of the
external line) of the graph. This is a straightforward generalization of the
temporal gauge in the previous section.
We will use the particle field theory case as a toy model and do the things
we do for string theory, for this simpler example. Quite surprisingly,
quantization in the temporal gauge coincides with the stochastic quantization!

Using this time coordinate, the
Feynmann graph can be reorganized as in Fig.3.
Let us construct the Hamiltonian describing the evolution of particles
in this time coordinate frame. Since we should deal with the processes
involving creations and annihilations of particles, we introduce the
creation $\phi^\dagger (x)$ and the annihilation $\phi (x)$ operators
satisfying
\eq
[\phi (x),\phi^\dagger (y)]=\delta^4 (x-y).
\en
Notice that this commutation relation is quite different from the usual
equal-time commutation relation of the field theory.
As is seen from Fig.3, the Hamiltonian includes the terms
\begin{enumerate}
\item kinetic term (Fig.4a),
\item four point vertex (Fig.4b),
\item two particle annihilation (Fig.4c).
\end{enumerate}
Notice that the four point vertex of the types depicted in Fig.4d can also
occur. However such kinds of vertices appear in very special configurations
when
the point at the vertex has more than two nearest boundaries. Since such
configurations has a vanishing measure in the whole configurations of the
Feynmann graphs, we can omit such terms from the Hamiltonian.

Each of the three terms can be expressed by $\phi$ and $\phi^\dagger$ as
$\phi^\dagger (-\partial_\mu\partial^\mu +m^2)\phi$ (kinetic term),
$(\phi^\dagger )^3\phi$ (four point vertex), and $\phi^2$ (two particle
annihilation). Carefully enumerating the symmetry factors, one can show that
the Hamiltonian is
\eq
H=\frac{1}{2}\int dx^4(\phi^2(x) +
\frac{\delta S}{\delta \varphi (x)}[\phi^\dagger ]
\phi (x)).
\label{pH}
\en
This form of Hamiltonian is valid for any action $S[\varphi ]$.

Feynmann graphs with $n$ external lines can be expressed by using this
Hamiltonian as follows. Let us prepare the bra and ket vacua satisfying
\eq
\phi (x)|0\rangle =\langle 0|\phi^\dagger (x)=0
\en
and consider the quantity
\eq
\langle 0|e^{-tH}\phi^\dagger (x_1)\cdots \phi^\dagger (x_n)|0\rangle .
\en
This represents the amplitude corresponding to the processes where
$n$ particles evolve into nothing during the period $t$. Since the
Hamiltonian in eq.(\ref{pH}) annihilates the vacuum, this means that
the particles vanish before the time $t$. Therefore $n$ particle amplitude
should be represented as
\eq
\lim_{t\rightarrow \infty}
\langle 0|e^{-tH}\phi^\dagger (x_1)\cdots \phi^\dagger (x_n)|0\rangle .
\label{npart}
\en

Thus Feynmann amplitudes with $n$ external lines are represented in a
form which looks not so familiar. Moreover the Hamiltonian in
eq.(\ref{pH}) does not have a form which becomes hermitian when it is Wick
rotated. The formulation of the field theory we obtained here is
completely different from the usual ones.
Actually this way of expressing
the amplitudes is closely related to the stochastic quantization\cite{pw}.
The Hamiltonian in eq.(\ref{pH}) corresponds to the kernel of the
Fokker-Planck equation in the stochastic quantization. Indeed, if one
defines the probability distribution $P(t;\varphi (x))$ as
\eq
P(t;\varphi (x))=
\langle 0|e^{-tH}\prod_{x}\delta (\phi^\dagger (x)-\varphi (x))|0\rangle ,
\en
one can deduce the equation
\eq
\partial_tP(t;\varphi (x))
=
-\frac{1}{2}\int dx^4(\frac{\delta^2}{\delta \varphi (x)\delta \varphi (x)}
+\frac{\delta}{\delta \varphi(x)}\frac{\delta S}{\delta \varphi (x)})
P(t;\varphi (x)),
\label{FP}
\en
which is exactly the Fokker-Planck equation. In the limit $t\rightarrow
\infty$,
$P$ approaches the stationary solution $e^{-S}$, which recovers the
path integral weight. Thus the time coordinate $t$ in the temporal gauge
corresponds to the fictitious time in the stochastic quantization.

There is a more direct way of understanding the relation between the
field theory in the temporal gauge and the stochastic quantization.
Let us consider a Feynmann graph reorganized according to the temporal
gauge time coordinate (for example, Fig.3). One can cut the graph into
pieces by cutting it at every two particle annihilation vertex. Then one
obtains tree graphs (Fig.5). Conversely if
one assigns the sum of tree graphs with sources to each external line and
connects the sources two by two, one recovers all the Feynmann graphs.
The sum of the tree graphs yields the solution to the equation
\eq
\partial_t\varphi (t,x)= -\frac{\delta S}{\delta \varphi (x,t)}+\eta (x,t),
\en
with the source $\eta (t,x)$ and the rule of connecting the sources can be
interpreted as $\eta $ is a Gaussian noise,
\eq
\langle \eta (t,x)\eta (t^\prime ,y)\rangle =\delta (t-t^\prime )\delta (x-y),
\en
which is exactly the formalism of the stochastic quantization.

In this formulation of field theory, the S-D equations for
$n$-particle amplitudes can readily be derived.
As in the same way as we obtained eq.(\ref{FP}) for the probability
distribution
$P(t;\varphi (x))$, we can derive the following equation:
\eq
\partial_t
\langle 0|e^{-tH}\phi^\dagger (x_1)\cdots \phi^\dagger (x_n)|0\rangle
=
-\langle 0|e^{-tH}[H,\phi^\dagger (x_1)\cdots \phi^\dagger (x_n)]|0\rangle .
\en
In the limit $t\rightarrow \infty$,
the quantity
$\langle 0|e^{-tH}\phi^\dagger (x_1)\cdots \phi^\dagger (x_n)|0\rangle $
approaches the stationary solution of the above equation and we obtain
\eq
\lim_{t\rightarrow \infty}
\langle 0|e^{-tH}[H,\phi^\dagger_1 (x_1)\cdots \phi^\dagger_n (x_n)]|0\rangle
=0.
\en
This gives exactly the S-D equation for $n$-particle amplitude.

Thus we have seen that we can quantize a particle field theory in the temporal
gauge and obtain the stochastic quantization of it. The string
field theories in the temporal gauge, which we will discuss in the following
sections have many features in common with the particle field theory in this
section.

\section{$c=0$ String Field Theory in the Temporal Gauge}
\hspace{5mm}
In this section, we will construct $c=0$ string field theory in the temporal
gauge. Let us proceed as in the previous section.
Feynmann graphs in string theory are two dimensional surfaces. We can
introduce the time coordinate as discussed in the introduction, and reorganize
the surfaces according to this time coordinate (Fig.6).

We would like to
construct the Hamiltonian describing the evolution of strings in this time
coordinate.  In the time evolution, following elementary processes occur.
\begin{enumerate}
\item A string propagates. (Fig.7a)
\item String interactions. (Fig.7b)
\item A string disappears. (Fig.7c)
\end{enumerate}
A peculiar feature of the temporal gauge string theory is the third process.
Notice that the inverse of such process in which a string appears from nothing
cannot occur because of the definition of the time coordinate.

We would like to construct a Hamiltonian $H$ representing
such processes for $c=0$ string theory.
As in the case of the particle field theory, we need the creation
and the annihilation operators of strings. A string in
$c=0$ string theory is
labelled by its length.
Let $\Psi^{\dagger} (l)$
($\Psi (l)$) be the creation (annihilation) operator which satisfies
\eq
\mbox{[} \Psi (l), \Psi^{\dagger }(l') \mbox{]}=\delta (l-l').
\label{comm1}
\en
These operators create or annihilate a string with length $l$.
To be precise, $\Psi^\dagger$ creates a string with one marked point.
They act on the Hilbert space generated from the vacuum
$|0>$ and $<0|$:
\eqs
\Psi (l)|0>=<0|\Psi^{\dagger} (l)=0.
\ens

The processes above would be expressed by the terms of
the form $(\Psi^{\dagger} )^n(\Psi )^m$ ($n=0,1,2,...$, $m=1,2,...$)
in $H$. As some of the four point vertices did not appear in the point particle
case, not all such terms appear in the Hamiltonian $H$.
A term which is ``simpler'' in the sense that it has less $n$ and $m$, is more
generic and is likely to appear in the Hamiltonian.
We can also use the dimensional analysis to figure out which
terms are present in $H$.
The scaling dimensions of the creation operator $\Psi^{\dagger}$ is the same
as the disk partition function as we will see later. For the $c=0$ string
case, $[\Psi^{\dagger}]=L^{-\frac{5}{2}}$\cite{KPZ},
where $L$ denotes the dimension of
the length $l$. From eq.(\ref{comm1}), the dimension of $\Psi $ becomes
$[\Psi ]=L^{\frac{5}{2}}$. The scaling dimension of the Hamiltonian $H$ is
the inverse of that of the time coordinate $D$
\footnote{In this section, we change the notation a bit. Here $D$ denotes the
time coordinate and $t$ denotes the cosmological constant. }
, which was evaluated in
\cite{KKMW} as $[D]=L^{\frac{1}{2}}$.

Let us see what kind of terms are possible for $H$. The kinetic term
(process 1.), has the form
\eq
\int_0^\infty dl\int_0^\infty dl^\prime
\Psi^\dagger (l)K(l,l^\prime ;t)\Psi (l^\prime ),
\en
where the kernel $K(l,l^\prime ;t)$ depends on the cosmological constant
$t$.
Here let us fix the form of $K(l,l^\prime ;t)$ on the following
plausible assumptions. Firstly we assume that $K(l,l^\prime ;t)$ has its
support at $l=l^\prime$ as a function of $l$. This means that the length of
the string changes only infinitesimally in an infinitesimal period of time.
The second assumption is that only positive integer powers of $t$ appear in
$K(l,l^\prime ;t)$. This implies that only surfaces of vanishing area are
involved in the elementary process. Actually there does not exist any function
$K(l,l^\prime ;t)$ satisfying the above conditions and with the right
dimension.
Therefore we conclude that our Hamiltonian has no kinetic term.

The tadpole term (process 3.), has the form
\eq
\int_0^\infty dl\rho (l;t)\Psi (l).
\en
Here we also assume $\rho (l;t)$ has its support at $l=0$, and only
positive integer powers of $t$ are involved. The reasons for these assumptions
are the same as above. The only possible form of $\rho$ is
\eq
\rho (l;t)=c_1\delta^{\prime \prime}(l)+c_2t\delta (l),
\en
where $c_1,~c_2$ are dimensionless constants.

The string interaction terms are of the form
\eq
\int_0^\infty dl_1\cdots \int_0^\infty dl_n
\int_0^\infty dl^\prime_1\cdots \int_0^\infty dl^\prime_m
\Psi^\dagger (l_1)\cdots \Psi^\dagger (l_n)
K_{n,m}(l_1,\cdots ,l_n;l^\prime_1,\cdots ,l^\prime_m)
\Psi (l^\prime_1)\cdots \Psi (l^\prime_m).
\en
Here we assume that $K_{n,m}$ $n\geq 2,~m\geq 1$ is proportional to
$\delta (l_1+\cdots l_n-l^\prime_1-\cdots l_m^\prime )$ which expresses the
conservation of the length in the interaction. This is reminiscent of the
three-Reggeon vertex. In the continuum limit of the dynamical triangulation,
this is the leading term of the interaction of this kind. Without any
dimensionful parameters except for the lengths, the simplest term which has
the right dimension $L^{-\frac{1}{2}}$ is
\eq
 \int_0^{\infty}dl_1\int_0^{\infty}dl_2
  \Psi^{\dagger}(l_1)\Psi^{\dagger}(l_2)\Psi (l_1+l_2)(l_1+l_2).
\en

If one allows the string coupling constant $g$ whose dimension is $L^{-5}$
\cite{DS}
to appear in the vertex, the vertex of the form
\eq
g\int_0^{\infty}dl_1\int_0^{\infty}dl_2
\Psi^\dagger (l_1+l_2)\Psi (l_1)\Psi (l_2)l_1l_2,
\en
becomes possible.

Putting all these terms together, we obtain the following Hamiltonian:
\eqa
H
&=&
\int_0^{\infty}dl_1\int_0^{\infty}dl_2
  \Psi^{\dagger}(l_1)\Psi^{\dagger}(l_2)\Psi (l_1+l_2)(l_1+l_2)
\nonumber
\\
& &
  +g\int_0^{\infty}dl_1\int_0^{\infty}dl_2
  \Psi^{\dagger}(l_1+l_2)\Psi (l_1)\Psi (l_2)l_1l_2
\nonumber
\\
& &
+\int _0^{\infty}dl(3\delta ''(l)-\frac{3}{4}t\delta (l))\Psi (l).
\label{pham}
\ena
We have fixed the normalization of each term and the form of $\rho$ by
rescaling $\Psi ,~l,~t,$ and the time variable $D$.

Thus a possible
Hamiltonian for $c=0$ string theory is obtained. Of course, the above
derivation
is not rigorous. We derived the Hamiltonian on many assumptions and only the
simplest interactions are included. What is remarkable is that the Hamiltonian
in eq.(\ref{pham}) really describes $c=0$ string theory. We will show this by
deriving the Virasoro constraints from this Hamiltonian in the following.

As was stressed in the introduction, the Hamiltonian in the temporal gauge is
in
close connection with the matrix model S-D equation. Therefore it is quite
natural that we can derive the Virasoro constraints from this Hamiltonian
if we have the right Hamiltonian. In order to do so, let us
first express string amplitudes in terms of this Hamiltonian. The way to do
so is the same as in the particle theory case. $n$ string amplitude can
be expressed as
\eq
\lim_{D\rightarrow \infty}
<0|e^{-D{\cal H}}\Psi^{\dagger }(l_1)\cdots \Psi^{\dagger }(l_n)|0>.
\label{nloop}
\en
Expanding perturbatively in terms of $g$, the contribution from the
connected surfaces
with $h$ handles and $b$ boundaries is proportional to $g^{-1+h+b}$.
The simplest case is the disk amplitude, which is obtained by setting $n=1$ and
$g=0$ in eq.(\ref{nloop}). The scaling dimension of $\Psi^\dagger$ thus
should coincide with that of the disk amplitude.

As in the derivation of the S-D equation in section 2, we obtain
the following equation for the string amplitudes above:
\eq
\lim_{D\rightarrow \infty}
<0|e^{-D{\cal H}}[H,\Psi^{\dagger }(l_1)\cdots \Psi^{\dagger }(l_n)]|0>=0.
\label{WD}
\en
In the point of view of 2D gravity, this equation means that the amplitudes
do not change if one acts the Hamiltonian operator on the boundaries. In other
words, this equation corresponds to the Wheeler-DeWitt equation in the temporal
gauge. Also as was discussed in section 1, this equation should be expressed as
an integral of matrix model S-D equation along the boundary loops.
Therefore we should be able to derive the Virasoro constraints from this
equation.

One can do so as follows. Let us prepare the generating functional of the
string amplitudes:
\eq
Z(J)=
\lim_{D\rightarrow \infty}
<0|e^{-D{\cal H}}e^{\int dlJ(l)\Psi^{\dagger }(l)}|0>.
\en
In terms of the generating functional of the connected amplitudes $\ln Z(J)$,
eq.(\ref{WD}) can be rewritten as
\eqa
& &
\int_0^{\infty }dlJ(l)\{
l\int_0^ldl'
\mbox{[}
\frac{\delta^2\ln Z(J)}{\delta J(l')\delta J(l-l')}
+\frac{\delta \ln Z(J)}{\delta J(l)}\frac{\delta \ln Z(J)}{\delta J(l-l')}
\mbox{]}
\nonumber
\\
& &
\hspace{2cm}
+gl\int_0^{\infty }dl'J(l')l'
\frac{\delta \ln Z(J)}{\delta J(l+l')}
\nonumber
\\
& &
\hspace{2cm}
+\rho (l)\} =0.
\label{eqc}
\ena
If one functionally differentiates eq.(\ref{eqc}) by $J$ and puts $J=0$ later,
one recovers eq.(\ref{WD}). As we stated in the above paragraph, eq.(\ref{WD})
asserts that the partition function does not change under the simultaneous
evolution of all the boundaries. On the other hand, the matrix model
S-D equation corresponds to a deformation of only one boundary.
It is easy to see that the evolution of one boundary corresponds to the
quantity between $\{$ and $\}$ in eq.(\ref{eqc}). If it vanishes, i.e.
\eqa
& &
l\int_0^ldl'
\mbox{[}
\frac{\delta^2\ln Z(J)}{\delta J(l')\delta J(l-l')}
+\frac{\delta \ln Z(J)}{\delta J(l)}\frac{\delta \ln Z(J)}{\delta J(l-l')}
\mbox{]}
\nonumber
\\
& &
\hspace{2cm}
+gl\int_0^{\infty }dl'J(l')l'
\frac{\delta \ln Z(J)}{\delta J(l+l')}
\nonumber
\\
& &
\hspace{2cm}
+\rho (l)=0,
\label{mmsdl}
\ena
eq.(\ref{eqc}) is satisfied. Conversely, as will be explained
in the next paragraph,
eq.(\ref{eqc}) implies eq.(\ref{mmsdl}) on some condition.
Eq.(\ref{mmsdl}) should be the integral of the matrix model S-D
equation along the boundary. In the case at hand,
the boundary is homogeneous and
the integral means multiplication by the length $l$ of the boundary. Hence
eq.(\ref{mmsdl}) divided by $l$ should be the matrix model S-D equation.
In the rest of this section we will show that eq.(\ref{mmsdl}) certainly
yields the Virasoro constraints.

Before doing so, let us argue that eq.(\ref{eqc}) implies eq.(\ref{mmsdl}).
Eq.(\ref{eqc}) can be regarded as the
generating functional of S-D equations.
The connected loop amplitudes can be expanded in terms
of $g$ as
\eq
\frac{\delta^n \ln Z(J)}{\delta J(l_1)\cdots \delta J(l_n)}|_{J=0}
=
\sum_{h=0}^{\infty }g^{n-1+h}<w(l_1)\cdots w(l_n)>_h.
\label{pert}
\en
Here $<w(l_1)\cdots w(l_n)>_h$ is the contribution from the surfaces with
$h$ handles.
Eq.(\ref{eqc}) can be reduced to equations for $<w(l_1)\cdots w(l_n)>_h$
which can be solved inductively
starting from the amplitudes with fewer loops and handles.
In order to solve each of these equations, we should impose an appropriate
boundary condition. Here we will require that $<w(l_1)\cdots w(l_n)>$
vanishes when any of $l_i$ goes to infinity. Considering the loop amplitudes
as the wave function of two dimensional quantum gravity corresponding to a
multi-loop (or multi-universe) state, it is natural to impose such a
condition \cite{MSS}.
It is possible to
show that if the equation of the form eq.(\ref{eqc}) has a solution
$<w(l_1)\cdots w(l_n)>_h$ satisfying such a boundary
condition, the solution is unique.
Therefore if eq.(\ref{mmsdl}) has a solution satisfying the boundary condition,
it should coincide with the unique solution of eq.(\ref{eqc}). We will show
that eq.(\ref{mmsdl}) is equivalent to the Virasoro constraints and the
the loop amplitudes of $c=0$ string theory provide
 the solution to eq.(\ref{mmsdl}) satisfying the boundary condition. Thus
eq.(\ref{eqc}) implies eq.(\ref{mmsdl}) on such a boundary
condition.

Now let us derive the Virasoro constraints from eq.(\ref{mmsdl}).
In \cite{FKN}, the authors transform the loop equation into the relations
between the correlation functions of the local operators ${\cal O}_n$,
which appear if one expands the macroscopic loop operator
$w(l)$ in terms of $l$:
\eq
w(l)=g\sum_{n\geq 0}\frac{l^{n+1/2}}{\Gamma (n+\frac{3}{2})}{\cal O}_n.
\label{local}
\en
The factor $g$ on the right hand side is put so that the insertions of
local operators do not change the order in $g$.
In order to derive equations for amplitudes with insertions of such local
operators, we should choose the source $J(l)$ so that
\eq
\int_0^{\infty}dlJ(l)l^{n+\frac{1}{2}}=g^{-1}\Gamma (n+\frac{3}{2})\mu_n.
\label{source}
\en
Then the generating
functional $\ln Z(J)$ can be considered as the generating functional
of connected correlation functions of the local operators:
\eqs
\ln Z(J)=<e^{\sum \mu_n{\cal O}_n}>.
\ens

Therefore substituting eq.(\ref{source}) into eq.(\ref{mmsdl}), we can obtain
relations between the correlation functions of the local operators. One thing
one should notice in doing so is that the loop operator $w(l)$ cannot
always be expanded as eq.(\ref{local}).
Since the amplitudes for the disk and the cylinder have a special part which
cannot be written as eq.(\ref{local}),
we have the following ansatz for the solution of
eq.(\ref{mmsdl}):
\eqas
\frac{\delta \ln Z(J)}{\delta J(l)}
=
\frac{l^{-\frac{5}{2}}}{\Gamma (-\frac{3}{2})}
-\frac{3t}{8}\frac{l^{-\frac{1}{2}}}{\Gamma (\frac{1}{2})}
+\frac{g}{2\pi}\int_0^{\infty}dl'J(l')\frac{\sqrt{ll'}}{l+l'}
+g\sum_{n\geq 0}\frac{l^{n+1/2}}{\Gamma (n+\frac{3}{2})}
\frac{\delta \ln Z(J)}{\delta \mu_n}.
\enas
Substituting this into eq.(\ref{mmsdl}),
we obtain the following infinite number of equations.
\eqa
& &
2\frac{\partial Z}{\partial \mu_1}
=
-\frac{1}{g}(\frac{3t}{8}-\frac{\mu_0}{2})^2Z
-\sum_{n=1}^{\infty}(n+\frac{1}{2})\mu_n\frac{\partial Z}{\partial \mu_{n-1}},
\\
& &
2(\frac{\partial Z}{\partial \mu_2}
-\frac{3t}{8}\frac{\partial Z}{\partial \mu_0})
=
-\frac{1}{16}Z
-\sum_{n=0}^{\infty}(n+\frac{1}{2})\mu_n\frac{\partial Z}{\partial \mu_{n}},
\\
& &
2(\frac{\partial Z}{\partial \mu_{p+3}}
-\frac{3t}{8}\frac{\partial Z}{\partial \mu_{p+1}})
=
-g\sum_{n=0}^p\frac{\partial^2 Z}{\partial \mu_n\mu_{p-n}}
-\sum_{n=0}^{\infty}(n+\frac{1}{2})\mu_n\frac{\partial Z}{\partial \mu_{n+p+1}}
\\
& &
\hspace{7cm}(p\geq 0).
\nonumber
\ena
These equations coincide with the Virasoro constraints in \cite{FKN} up to
some rescalings of parameters.
Hence the Hamiltonian we constructed yields the Virasoro constraints.
The loop amplitudes of two dimensional
gravity provide the unique solution of eq.(\ref{mmsdl}).

Thus the Hamiltonian we constructed on some assumptions really describes
$c=0$ string theory. As was discussed in section 1, the Hamiltonian in the
temporal gauge is closely related to the S-D equation of the matrix model.
We have checked that the matrix model S-D equation
can be derived from the Hamiltonian we proposed. Conversely, if one looks at
the matrix model S-D equation carefully, one should be able to construct
the string field Hamiltonian. Watabiki\cite{wata} did so for $c=0$ string
theory and reproduced our Hamiltonian. In the next section, we will construct
the string field Hamiltonian for $c\leq 1$ string theory in this way.

\section{$c\leq 1$ String Field Theory in the Temporal Gauge}
\hspace{5mm}
In this section, we will go on to construct the string field theory in the
temporal gauge for $c\leq 1$ string theory. Here we will concentrate on
$c=\frac{1}{2}$ case. Discussions on more general cases will be found
in \cite{IIKMNS}.

The strategy to construct the Hamiltonian is the one explained at the end
of the last section. Since the temporal gauge Hamiltonian is closely related
to the matrix model S-D equation, we will investigate the matrix model
S-D equation and infer the form of the Hamiltonian. $c=\frac{1}{2}$ string
theory can be realized by putting the Ising spins on the worldsheet.
In the matrix model approach, Ising spins are introduced on the worldsheet
by considering the two matrix model. The S-D
equations of the two matrix model were studied in \cite{GN} and the authors
derived the $W_3$ constraints from them. We will first consider the continuum
limits of Gava-Narain's S-D equations and then construct the Hamiltonian
from them. Thus our Hamiltonian is made to yield the $W_3$ constraints and
it is clear that it describes $c=\frac{1}{2}$ string theory.

When there exist spin degrees of freedom on the worldsheet, a state of a string
is labeled by its length and the spin configuration on it.
In the continuum limit, an Ising spin configuration
may be represented by a state of $c=\frac{1}{2}$ conformal field theory (CFT).
Let us define an operator $w(l;|v\rangle )$ representing a string state with
length $l$ and the spin configuration corresponding to $|v\rangle $ which
is a state of $c=\frac{1}{2}$ CFT. We will denote $n$-string amplitude by
\eq
<w(l_1;|v_1\rangle )w(l_2;|v_2\rangle )\cdots w(l_n;|v_n\rangle )>.
\label{ampl}
\en
The matrix model S-D equations give relations among such string amplitudes.

Let us sketch how Gava and Narain obtained the $W_3$ constraints from the
matrix model S-D equations. The $W_3$ constraints are expected to come
from equations about the loop amplitudes in which the Ising spins on all the
boundary loops are, say, up. Suppose the partition function of the
dynamically triangulated surfaces with boundaries on which all the Ising spins
are up. If one takes one triangle from a boundary, the following three things
can happen. (Fig.8 )
\begin{enumerate}
\item The boundary loop splits into two.
\item The boundary loop absorbs another boundary.
\item The spin configuration on the boundary loop changes.
\end{enumerate}

The matrix model S-D equation is a sum of three kinds of terms
corresponding to the above processes.
In the first and the second process, only boundaries with all the spins up
can appear. The third process is due to the matrix model action.
A boundary loop on which one spin is down and all the others are up can
appear in this process. In order to derive the $W_3$ constraints, one should
somehow cope with this mixed spin configuration. Gava and Narain then
considered the loop amplitudes with one loop having such a spin configuration
and all the other loops having all the spins up.
They obtained two S-D equations corresponding to the processes of
taking away the triangle attached to the link on which the Ising spin is down
and the one attached to the next link.
Those equations also consist of the terms corresponding to the above three
processes.
With these two equations, one can
express the loop amplitude with one mixed spin loop insertion by loop
amplitudes
with all the spins up. Thus they can obtain closed equations for loop
amplitudes
with all the spins up and the $W_3$ constraints were derived from them.

Now let us construct the continuum limit of Gava-Narain's equations.
The matrix model S-D equation describes the change of the amplitude
eq.(\ref{ampl}) , when one takes a triangle away from a boundary.
The continuum version of it should describe what happens when one
deforms the amplitude eq.(\ref{ampl}) at a point on a boundary.
In principle, by closely looking at the discrete S-D equations and taking the
continuum limit, one should be able to figure out what the continuum S-D
equations will be. However, in actuality, it is not an easy task, because of
the existence of the non-universal parts in the loop operators and the operator
mixing between various loop operators. Therefore, here we will construct
the continuum S-D equations by assuming some properties of them and check the
validity of the assumptions later by deriving the $W_3$ constraints from
them.

First we will assume that the continuum S-D equations consists of the three
terms corresponding to the three processes in the above paragraph (cf. Fig.8).
In the derivation of the $W_3$ constraints, Gava and Narain
started from loops with
all the spins up. Such a spin configuration was represented as a state
of $c=\frac{1}{2}$ CFT in \cite{ni}\cite{cardy}.
Let us denote such a state by $|+\rangle $.
It is clear that if such a loop splits into two, it results in
two loops with all the spins up. Also if a loop with all the spins up absorbs
another one, we obtain another loop with all the spins up. Therefore
the process of splitting and merging is particularly simple for such kind of
loops. The first S-D equation Gava and Narain considered corresponds to the
deformation of the loop amplitude eq.(\ref{ampl}) with
$|v_1\rangle =|v_2\rangle
=\cdots =|v_n\rangle =|+\rangle $. The equation in the continuum limit should
be
\eqa
& &
\int_0^l dl'<w(l';|+\rangle )w(l-l';|+\rangle )
w(l_1;|+\rangle )\cdots w(l_n;|+\rangle )>
\nonumber
\\
& &
+g\sum_kl_k<w(l+l_k;|+\rangle )w(l_1;|+\rangle )\cdots w(l_{k-1};|+\rangle )
w(l_{k+1};|+\rangle )\cdots w(l_n;|+\rangle )>
\nonumber
\\
& &
+<w(l;{\cal H}(\sigma )|+\rangle )w(l_1;|+\rangle )\cdots w(l_n;|+\rangle )>
\approx 0.
\label{eqone}
\ena
Here the first term corresponds to the process 1 in the above and the second
term is for the process 2. The string coupling constant $g$ comes in front
of the second term as in the case of $c=0$ string.
The last term describes the process 3. We have assumed that the local change
of the spin configuration in S-D equation can be represented by the
operator ${\cal H}(\sigma )$ acting on the states of $c=\frac{1}{2}$ CFT.
$0\leq \sigma < 2\pi$ is the coordinate of the point where
the local change occurs. The coordinate $\sigma $ on the loop is taken so that
the induced metric on the loop becomes independent of $\sigma $.
$\sigma =0$ is taken to be the marked point of the loop.
$\approx 0$ here means that
as a function of $l$, the quantity has its support at $l=0$. Therefore the
left hand side of eq.(\ref{eqone})
is equal to a sum of derivatives of $\delta (l)$.
These delta functions correspond to processes in which a string with
vanishing length disappears. In the point of view of string field theory,
such processes are expressed by the tadpole terms.

The two other equations which Gava and Narain used were obtained by taking a
triangle away from
$w(l;{\cal H}(\sigma )|+\rangle )$. The triangles to be considered were
the one attached to the link where ${\cal H}(\sigma )$ is inserted and the
one next to it.
In the continuum limit, these correspond to deforming the loop
$w(l;{\cal H}(\sigma )|+\rangle )$ at the point $\sigma$.
One should take a limiting procedure to obtain such an equation rigorously.
\footnote{
Consider a loop
$w(l;{\cal H}(\sigma )|+\rangle )$ and deform at a point near $\sigma $ and
take the limit in which the point tends to $\sigma $}
The S-D equation becomes
\eqa
& &
\int_0^l dl'<w(l';|+\rangle )w(l-l';{\cal H}(\sigma )|+\rangle )
w(l_1;|+\rangle )\cdots w(l_n;|+\rangle )>
\nonumber
\\
& &
+g\sum_kl_k<w(l+l_k;{\cal H}(\sigma )|+\rangle )
w(l_1;|+\rangle )\cdots w(l_{k-1};|+\rangle )
w(l_{k+1};|+\rangle )\cdots w(l_n;|+\rangle )>
\nonumber
\\
& &
+<w(l;({\cal H}(\sigma ))^2|+\rangle )w(l_1;|+\rangle )\cdots w(l_n;|+\rangle
)>
\approx 0.
\label{eqtwo}
\ena
Here we have a new kind of loop $w(l;({\cal H}(\sigma ))^2|+\rangle )$, with
two ${\cal H}$ insertions. Actually the term involving this kind of loop
vanishes. Indeed it is possible to show that the combination of the two
equations eqs.(\ref{eqone})(\ref{eqtwo}) with the last term of eq.(\ref{eqtwo})
omitted yield the $W_3$ constraints. We will not present here the derivation
of the $W_3$ constraints, which will be found in \cite{IIKMNS}.

Thus we have seen that assuming that amplitudes involving
$w(l;({\cal H}(\sigma ))^2|+\rangle )$ vanish $W_3$ constraints are
derived from the S-D equations. In \cite{IIKMNS}, this assumption was
proved by using another S-D equation. Here let us give a different
account of this assumption. In the following we will see that the
nature of the operator ${\cal H}(\sigma )$ seems to be responsible
for $({\cal H}(\sigma ))^2|+\rangle $ to be zero.

At the discrete level,
acting on the state $|+\rangle $ corresponding to a loop with all the spins
up, this operator is supposed to insert one down spin.
Let us express such an operator at the continuum level.
$c=\frac{1}{2}$ CFT can be realized as the field theory of left and right free
Majorana fermions, $\psi$ and $\bar{\psi}$.
Indeed, one can construct a Fock space of a free fermion starting
from the state $|+\rangle $ as follows.
We will take $|+\rangle $ as the vacuum.
In a general spin configuration on a loop, up and down spins coexist.
We can consider such a configuration as a loop through which the domain walls
of the Ising spin penetrate. The domain walls are self-avoiding walks
on the worldsheet and we will express them using a fermionic variable.
Let us introduce $\chi (\sigma )$ and $\chi^\dagger (\sigma )$ on the loop
$0\leq \sigma < 2\pi$, satisfying
\eq
\{ \chi (\sigma ), \chi^\dagger (\sigma ')\} =\delta (\sigma -\sigma ').
\en
$\chi^\dagger $ is the creation operator of a domain wall.
If a domain wall is penetrating through a loop at the position $\sigma $,
we express it by acting $\chi^\dagger (\sigma )$ on the state. $\chi $ is
the annihilation operator. Since there are no domain walls penetrating through
$|+\rangle $,
\eq
\chi (\sigma )|+\rangle =0.
\en
Thus any spin configurations on a loop can be expressed as a state with
an even number of $\chi^\dagger (\sigma )$'s acting on $|+\rangle $.

This fermionic Fock space is written in terms of the usual Fock space of the
free Majorana fermion $\psi $ as follows. $|+\rangle $ is
identified as a state of $c=\frac{1}{2}$ CFT in \cite{cardy}.
The identification is consistent with the equation
\eq
(\psi (\sigma )+i\bar{\psi}(\sigma ))|+\rangle =0.
\en
Therefore, $\chi $ and $\chi^\dagger $ can be identified as
\eqa
& &
\chi (\sigma )=\psi (\sigma )+i\bar{\psi}(\sigma )
\nonumber
\\
& &
\chi^\dagger (\sigma )=\psi (\sigma )-i\bar{\psi}(\sigma ).
\ena
Actually $|+\rangle $ is not in the Fock space of $\psi $, because norm
of it diverges including higher and higher modes of $\psi$. However,
by modifying the definition of the inner product \cite{cardy}, it is possible
to construct the Fock space from $|+\rangle $. We obtain a Fock space of
nonrelativistic fermions.

As was explained in the previous section,
an important effect of ${\cal H}(\sigma )$ is creating a pair of
domain walls very close to each other. In the leading order, such a pair on
$|+\rangle $ may be identified with
\eq
\chi^\dagger \partial \chi^\dagger (\sigma )|+\rangle .
\label{state}
\en
If one
inserts another ${\cal H}$ on the state in eq.(\ref{state}) close to $\sigma $,
it creates another pair of domain walls and we obtain a state
\eq
\chi^\dagger \partial \chi^\dagger (\sigma ')
\chi^\dagger \partial \chi^\dagger (\sigma )|+\rangle .
\label{hhstate}
\en
Since the states in eqs.(\ref{state})(\ref{hhstate}) have divergent norms,
they are not well-defined states even if we use the modified norm.
However it is plausible that the state in eq.(\ref{hhstate}) vanishes
in the limit $\sigma '\longrightarrow \sigma $ because of the fermi statistics.
Thus $w(l;({\cal H}(\sigma ))^2|+\rangle )=0$ may be proved.
We have not figured out how to make these arguments more rigorous.

Therefore ${\cal H}(\sigma )$ acting on $|+\rangle $ has the same effect
as that of an operator proportional to
$\chi^\dagger \partial \chi^\dagger (\sigma )$.
Considering that ${\cal H}(\sigma )$ induces the changes in the spin
configuration, which occur in the time evolution,
the most plausible of such operators is the Hamiltonian operator for the
$c=\frac{1}{2}$ CFT:
\eqa
{\cal H}(\sigma )
&=&
T(\sigma )+\bar{T}(\sigma )
\nonumber
\\
&=&
i\chi^\dagger \partial \chi^\dagger -
i\chi \partial \chi .
\ena

Thus we have seen that the S-D equations eqs.(\ref{eqone})(\ref{eqtwo}) with
the identification of the operator ${\cal H}(\sigma )$ as above give the
$W_3$ constraints of $c=\frac{1}{2}$ string theory. A straightforward
generalization of the S-D equations to more general states will be
as follows:
\eqa
\int_0^l dl'
& &
\sum_{|v^\prime \rangle ,|v^{\prime \prime }\rangle ,
{}~|v^\prime \rangle_{l'}*|v^{\prime \prime }\rangle_{l-l'}=|v\rangle_l}
<w(l';|v^\prime \rangle )w(l-l';|v^{\prime \prime }\rangle )
w(l_1;|v_1\rangle )\cdots w(l_n;|v_n\rangle )>
\nonumber
\\
& &
+g
\sum_kl_k\int_0^{2\pi }d\sigma^\prime
<w(l+l_k;|v\rangle_l *(e^{i\sigma^\prime {\cal P}}|v_k\rangle_{l_k}))
\nonumber
\\
& &
{}~~~~~~\times
w(l_1;|v_1\rangle )\cdots
w(l_{k-1};|v_{k-1}\rangle )
w(l_{k+1};|v_{k+1}\rangle )\cdots w(l_n;|v_n\rangle )>
\nonumber
\\
& &
+<w(l;{\cal H}(\sigma )|v\rangle )w(l_1;|v_1\rangle )\cdots w(l_n;|v_n\rangle
)>
\approx 0.
\label{sdeqg}
\ena
Again we have assumed that the S-D equation includes three processes in
Fig. 8.
Here we have introduced the product $*$ so that
\eq
|v_1\rangle_{l_1}*|v_2\rangle_{l_2},
\en
represents a loop made by merging the two loops $|v_1\rangle_{l_1}$ and
$|v_2\rangle_{l_2}$ at the marked points, with the spin
configuration inherited from them (Fig.9). We have assumed that when a string
splits into two or two strings merge, the process is described by such
product. The local change of the spin configuration on the string is again
described by the operator ${\cal H}(\sigma )$.

{}From such an S-D equation, one can obtain the string field Hamiltonian.
Let $\Psi (l;|v\rangle )$
($\Psi^\dagger (l;|v\rangle )$) denotes the annihilation (creation) operator
of a string with length $l$ and the spin configuration $|v\rangle $ satisfying
\eq
[\Psi (l;|v\rangle ), \Psi^\dagger (l';|v'\rangle )]
=
l\int_0^{2\pi} d\sigma \langle v'|e^{i\sigma {\cal P}}|v\rangle \delta (l-l').
\label{comm2}
\en
Namely the commutator of $\Psi (l;|v\rangle )$ and
$\Psi^\dagger (l';|v'\rangle )$ is nonzero only when $l=l'$ and $|v\rangle $
coincides with $|v'\rangle $ up to rotation.
The string field Hamiltonian can be obtained from eq.(\ref{sdeqg}) as
\eqa
H=
& &
\sum_{|v_i \rangle }\int_0^\infty dl_1\int_0^\infty dl_2
\Psi^\dagger (l_1;|v_1 \rangle )\Psi^\dagger (l_2;|v_2\rangle )
\Psi (l_1+l_2;|v_1\rangle_{l_1}*|v_2\rangle_{l_2} )
\nonumber
\\
& &
+g
\sum_{|v_i \rangle }\int_0^\infty dl_1\int_0^\infty dl_2
\Psi^\dagger (l_1+l_2;|v_1\rangle_{l_1}*|v_2\rangle_{l_2} )
\Psi (l_1;|v_1 \rangle )\Psi (l_2;|v_2\rangle )
\nonumber
\\
& &
+\sum_{|v\rangle }\int_0^\infty dl
\Psi^\dagger (l;{\cal H}(0)|v\rangle )\Psi (l;|v\rangle )
\nonumber
\\
& &
+\sum_{|v\rangle }\int_0^\infty dl\rho (l;|v\rangle )\Psi (l;|v\rangle ).
\label{hamleq}
\ena
Here $\rho (l;|v\rangle )$ expresses the tadpole term and it has its support
at $l=0$.

\section{Conclusions and Discussions}
\hspace{5mm}
In this note, we have reviewed string field theory in the temporal gauge.
We have shown how to construct the string field Hamiltonian for $c\leq 1$
string theory in the temporal gauge. The temporal gauge is peculiar in
many respects. For example, in this gauge,
even a disk amplitude is expressed as a sum of infinitely many
processes involving innumerable splitting of strings. It forms a striking
contrast to the case of the conformal gauge. The amplitudes can be calculated
by using the S-D equations of the string field.

For $c\neq 0$ case, we constructed the temporal gauge
string field Hamiltonian from the S-D equations. The Hamiltonian looks
quite similar to the Hamiltonian of the light-cone gauge string field theory
\cite{SFT}. It involves only three string interactions and the kinetic term
seems to be identified with $L_0+\bar{L}_0$.
Since the form of
the Hamiltonian is almost the same for any $c$, it might be possible to
construct the temporal gauge Hamiltonian in the same way
for $c>1$ case, especially for the critical string. This will be left
to the future investigations.

In the particle field theory case, the temporal gauge formulation yields
the stochastic quantization of the theory. Since the formalism in this
section is quite similar to the one in section 2, we suspect that
our string field theory may correspond to a stochastic quantization of
some theory. In this respect, the results of Jevicki and Rodrigues are
very suggestive. They considered the stochastic quantization of the
one-matrix model and by changing the variables from the matrix element to
the loop variable, they obtained the Hamiltonian in eq.(\ref{pham}) in the
continuum limit. In the continuum limit, the matrix element may correspond to
an infinitesimal line element of string. By changing the variable from the
string field $\Psi (l)$ to something representing the line element, we
may be able to obtain a string field theory which is more like the conventional
formulation of field theory.

\section*{Acknowledgements}
\hspace{5mm}
We would like to thank M.Fukuma, N.Kawamoto, M.Ninomiya, J.Nishimura,
Y.Okamoto, N.Tsuda, Y.Watabiki
and T.Yukawa for useful discussions and comments. N.I. would like to thank
organizers of the workshop for giving him a chance to give a talk at
a very stimulating and enjoyable workshop.

\end{document}